\documentclass[a4paper,12pt]{article}
\usepackage{times,latexsym,amssymb,overcite}

\usepackage[dvips]{graphics,color}
\begin{document}

\title{\vspace{-4cm}\hfill \textmd{\normalsize Applied Physics Reports no.\ 2003--22}\\[1cm]
Adsorption of Methanol on Aluminum Oxide: A Density Functional Study}
\author{{\O}yvind Borck$^{a,b,}$\footnote{Corresponding author, e-mail: oyvind.borck@phys.ntnu.no}\,\, and Elsebeth Schr\"{o}der$^b$\\
\small{\textit{$^a$Department of Physics, Norwegian University of Science 
and Technology,}}\\\small{\textit{ NO-7034 Trondheim, Norway}}
\\\small{\textit{$^b$Department of Applied Physics, Chalmers University of 
Technology and G\"{o}teborg}}\\ 
\small{\textit{University, SE-412 96 G\"{o}teborg, Sweden}}}
\date{July 1, 2003} 
  \maketitle
\begin{abstract}
Theoretical calculations based on density functional theory have made significant 
contributions to our understanding of metal oxides, their surfaces, and the binding of 
molecules at these surfaces. In this paper we investigate the binding of 
methanol at the $\alpha$-Al$_2$O$_3$(0001) surface using first-principles density
functional theory. We calculate the molecular adsorption energy of methanol to be 
$E^g_\mathrm{ads}=1.03$ eV/molecule. Taking the methanol-methanol interaction into account, we
obtain the adsorption energy $E_\mathrm{ads}=1.01$ eV/molecule. Our calculations indicate
that methanol adsorbs chemically by donating electron charge from the methanol oxygen to
the surface aluminum. We find that the surface atomic structure changes upon adsorption,
most notably the spacing between the outermost Al and O layers changes from 0.11 {\AA} to
0.33 {\AA}. 
\end{abstract}

\section{Introduction}
Today, adhesively bonded and coated aluminum structures find a range of technological
applications. Understanding the factors affecting the adhesion 
properties of oxide covered aluminum surfaces is important for controlling the long term
performance and durability of these structures. One such factor
is the nature and strength of the interfacial bonds between the resins of the
adhesive or coating and the substrate.
Despite their technological relevance, the understanding of metal oxide surfaces in
general is at a 
far less advanced level than most other solid surfaces.\cite{henrich&cox,noguera}
Experimental studies face complications in preparing the nearly perfect, clean surfaces 
needed for atomic scale surface investigations, and by the sheer complexity of the crystal 
and electronic structures. Theoretical studies have been impeded by the limited experimental 
data available for guidelines and the complexity of the oxide structures.

In this paper we present a theoretical investigation of methanol (CH$_3$OH) adsorption on clean 
$\alpha$-Al$_2$O$_3$(0001). The choice of methanol and $\alpha$-Al$_2$O$_3$(0001) is motivated
by the desire to contribute to a more fundamental understanding of adhesion of organic coatings
or adhesives at oxide-covered aluminum surfaces. Because of the size and chemical complexity of 
the binders found in coatings and adhesives, the interpretation of the adhesion mechanism can be
difficult. One strategy to simplify the analysis is to study the interaction of smaller 
molecules (representative of the various components of the binder) with the 
surface.\cite{mmblg:ass133:270} Following this approach, we will address the binding properties 
of the different functional groups separately.    

Methanol is a good starting point for such studies as it is 
the smallest organic molecule with a hydroxyl group, and because it is well 
characterized both experimentally and theoretically.\cite{methanol_clean}
In addition to the thermodynamically stable $\alpha$ phase, Alumina (Al$_2$O$_3$) has 
a number of metastable bulk phases ($\gamma$, $\eta$, $\theta$, $\kappa$,$\ldots$), 
some being stable up to rather high temperatures.\cite{carlo&co&ref_therein,levin}
Unlike most of the metastable aluminas (the exceptions are the $\theta$ and $\kappa$ phases), 
the atomic structure of $\alpha$-Al$_2$O$_3$ and its (0001) surface is agreed upon,\cite{ahn,wang,ryl:} 
and therefore represents
an ideal starting point for theoretical studies of adsorption on alumina surfaces.
The adsorption of methanol has been studied experimentally and theoretically on
a range of metal oxides, e.g., Cr$_2$O$_3$, ZrO$_2$, and TiO$_2$,
but to our knowledge only one theoretical\cite{dgkmprw:jmst469:7} and few 
experimental\cite{far:ss277:337,ngt:jpcb102:6831} investigations of methanol 
adsorption on Al$_2$O$_3$ have appeared in the literature. 

Water, which also has an OH-group, has been shown\cite{hsca:jpcb:5527} to adsorb both
as a molecule and dissociatively on $\alpha$-Al$_2$O$_3$. It is therefore of interest to study
both molecular and dissociative adsorption of methanol. In the present study our focus is on
molecular adsorption on the clean and perfect (no steps or defects) (0001)-surface of
$\alpha$-Al$_2$O$_3$. The effect of preadsorbed hydroxyl groups and the dissociative
methanol adsorption will be subject of a future study. 

\section{Density Functional Theory Calculations}

\begin{figure}[b]
\begin{center}
\scalebox{0.7}{\includegraphics{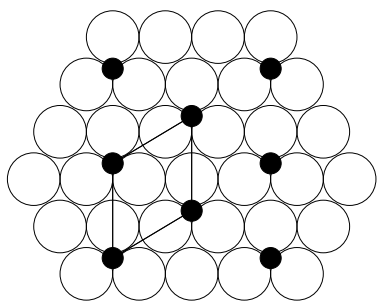}}
\caption{\label{metads}Schematic top view of the $\alpha$-Al$_2$O$_3$(0001) surface 
terminated by half a layer of aluminum. The open circles represent oxygen atoms, 
the full circles aluminum atoms. The 1 $\times$ 1 hexagonal surface unit cell is shown.}
\end{center}
\end{figure}

Bulk $\alpha$-Al$_2$O$_3$ has the corundum structure with alternating O and
Al layers along the [0001] direction with stacking sequence 
$\cdots$Al--O$_3$--Al--Al--O$_3$--Al$\cdots$. The (0001) surface is obtained
by cleaving the crystal between any of these layers. Only the surface obtained 
by cleaving between two aluminum layers is non-polar. Previous 
studies\cite{wang,hsca:jpcb:5527} have shown that in the absence of 
hydrogen and adsorbed H$_2$O this aluminum-terminated surface is 
more stable than the other (0001) surfaces. Figure~\ref{metads} shows a 
schematic top view of the aluminum terminated $\alpha$-Al$_2$O$_3$(0001) 
surface.

We here use first-principles, plane-wave density-functional theory\cite{dacapobok}
(DFT) to determine the adsorption (binding) energy and the equilibrium 
structure of methanol on $\alpha$-Al$_2$O$_3$(0001). 
The merit of first-principles calculations is that no
experimental input is needed for an accurate determination
of structural energies. We can thus compare calculations
based entirely on theory with experimental findings.

The DFT calculations were performed by employing
the DACAPO code\cite{dacapo} in the generalized gradient
approximation, using ultra-soft pseudopotentials and periodic boundary
conditions. The surface is represented by a slab of finite thickness
(8.3 {\AA}) alternating with a vacuum region (14.7 {\AA}).
The equilibrium geometries of the methanol molecule, the clean surface, 
and the adsorbate system are found by locally minimizing the Hellmann-Feynman 
forces until the sum of forces on the unrestrained atoms is less than 
0.05 eV/{\AA}. By adsorbing methanol on one side of the slab only an 
artificial dipole is created.
This artificial dipole is corrected for in a self-consistent manner.

The optimized geometry of the isolated (gas phase) methanol molecule is in excellent agreement 
with experiments, with bond lengths and bond angles deviating from the experimental 
values\cite{methanol_clean} by $\sim$~1~\%. The most important bond angle and
lengths for the present study are given in Table~\ref{tab:geometri}.
Upon optimization of the clean $\alpha$-Al$_2$O$_3$(0001) atomic structure, the 
inter-plane spacing is relaxed from the bulk spacing. The distance between the 
adjacent outermost aluminum and oxygen layer is reduced by 87.2~\%, in good agreement 
with previous theoretical work.\cite{wang,ryl:,hsca:jpcb:5527}

\section{Results and Discussion}

In our calculations we allow for molecular adsorption of one methanol molecule per surface
unit cell (Fig.~\ref{metads}), or one molecule per exposed Al atom.
Methanol adsorbs by the transfer of electronic charge from the O atom of methanol to 
create a bond to the exposed Al atom. 
Upon the adsorption of methanol, the layer spacing of $\alpha$-Al$_2$O$_3$(0001) changes 
markedly. We find that the spacing between the top Al and O layers changes from 
0.11 {\AA} to 0.33 {\AA}, whereas the lateral change of bond length between the surface O atoms
and the surface Al atoms is small (1--5\%). The changes in bond lengths within the adsorbed 
methanol molecule are modest (1--4 \%). The bond lengths of methanol in the gas phase and 
as adsorbed on the alumina surface are listed in Table~\ref{tab:geometri}.

The adsorption energy per methanol molecule is calculated from
\begin{equation}
  E_\mathrm{ads}^\mathrm{(g)}=-(E_\mathrm{tot}-E_\mathrm{s}-E_\mathrm{m}^\mathrm{(g)})\; ,
\end{equation}
where $E_\mathrm{tot}$ is the total energy of the system per unit cell,
$E_\mathrm{s}$ is the energy for the clean aluminum oxide surface per unit cell, and
$E_\mathrm{m}^\mathrm{(g)}$ is the energy of a methanol molecule away from the surface.
We calculate two adsorption energies for methanol, one for moving the
methanol molecules from the surface all the way into the gas phase ($E_\mathrm{m}^\mathrm{g}$),
and one for moving a complete layer of
methanol structurally unchanged off the surface ($E_\mathrm{m}$). 
The latter is dominated by 
the strength of the Al-O$_\mathrm{m}$ bond (`m' labels atoms in the molecule), 
and can very accurately be calculated by present-day DFT approximations. 
The $E_\mathrm{m}^\mathrm{(g)}$ adsorption  energy includes energetic contributions from the 
structural changes within the molecule, as well as  interactions between 
the methanol molecules (adsorbate-adsorbate interactions). 

We can compare our calculated adsorption energies with those of experimental
measurements. However, for all results of the experiments it should be
kept in mind that the chemisorption of methanol on Al$_2$O$_3$ is very sensitive to
surface defects and the presence of adsorbed water or hydrogen on the surface. 
In contrast, in our calculations we investigate a perfect (no defects or steps) clean
$\alpha$-Al$_2$O$_3$(0001) surface. From our DFT calculations 
we find that the energy of the Al-O$_\mathrm{m}$ bond is 
$E_\mathrm{ads}=1.01$ eV/molecule ($23.3 $ kcal/mol) and that 
$E_\mathrm{ads}^\mathrm{g}=1.03$ eV/molecule ($23.8 $ kcal/mol),
the small difference of 0.02 eV/molecule indicating the calculated effect of the 
molecule-molecule interactions and the structural changes in the methanol molecule. 
Given the difficulty in carrying out experimental measurements on methanol adsorption on
clean alumina surfaces, these theoretically obtained adsorption energies
agree reasonably well with the value 0.77 eV/molecule ($17.7$ kcal/mol)
obtained from temperature
programmed desorption experiments\cite{ngt:jpcb102:6831} 
at low coverages. 
In an UHV-chemisorption experiment on thin $\alpha$- and $\gamma$-Al$_2$O$_3$
films at low coverage an adsorption energy $\sim$ 0.3 eV/molecule (7~kcal/mol)
was found.\cite{far:ss277:337} 

\begin{figure}
\begin{center}
\scalebox{1}{\includegraphics{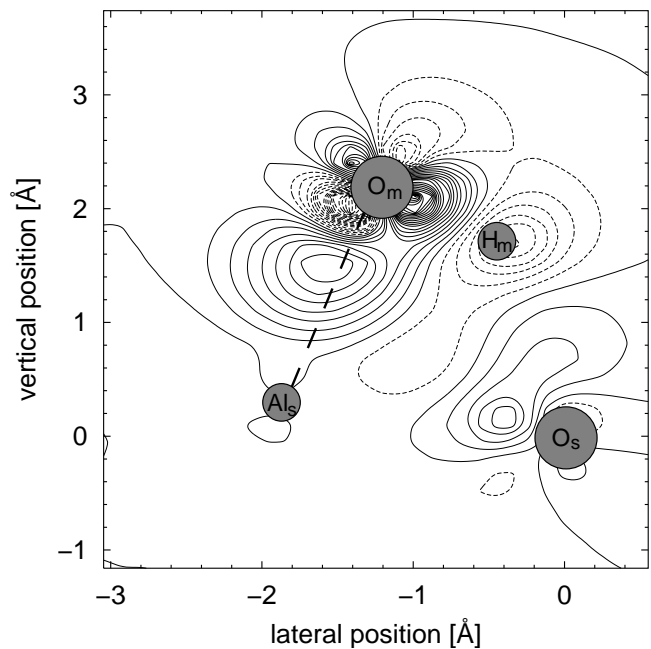}}
\caption{\label{chd}Contour plot of the electron density difference $\Delta n(\mathbf{r})$ in a cut 
perpendicular to the surface and passing through the surface Al (Al$_\mathrm{s}$)
and the methanol O (O$_\mathrm{m}$).
The atoms O$_\mathrm{s}$ and H$_\mathrm{m}$ are slightly out of plane.
The four atoms approximately in the 
plane are indicated by circles and named according to 
Table~\protect\ref{tab:geometri}. The bond between the methanol oxygen and the
surface aluminum is indicated by the long-dashed line.
The contour spacing is 0.02 electrons/{\AA}$^3$ with solid (dashed) lines for gain 
(loss) of negative charge.}
\end{center}
\end{figure}

Figure~\ref{chd} shows a contour plot of a cross section of the electron density  difference
$\Delta n(\mathbf{r})$ in a slice approximately through the outermost Al and O atoms
of the surface and the O and (hydroxyl-group) H atoms of the adsorbed methanol. 
$\Delta n(\mathbf{r})$ measures how electron charge is moved in space due
to the adsorbate-surface interaction. The electron density difference is defined as
$\Delta n(\mathbf{r})=n^\mathrm{m+s}(\mathbf{r})
-n^\mathrm{s}(\mathbf{r})-n^\mathrm{m}(\mathbf{r})$, where $n^\mathrm{m+s}(\mathbf{r})$
is the electron density of the full adsorbate system,
$n^\mathrm{s}(\mathbf{r})$ is the electron density of the surface without
methanol, and $n^\mathrm{m}(\mathbf{r})$ is that of methanol without the surface.
The atomic positions are in all cases keep fixed as for the full adsorbate system, 
thus $\Delta n(\mathbf{r})$ shows the change in electron density exclusively due
to the presence of the adsorbate.

Whereas an Al atom in the bulk of $\alpha$-Al$_2$O$_3$ is sixfold
coordinated, the surface Al atom (Al$_\mathrm{s}$) on the Al-terminated
(0001) surface is only three-fold coordinated, making this
site a strong electron acceptor. When methanol adsorbs,
the O atom of methanol (O$_\mathrm{m}$) binds to Al$_\mathrm{s}$ by donating 
electron density to creation of the Al$_\mathrm{s}$-O$_\mathrm{m}$ bond, forming 
a dative (covalent) bond. 
This is seen in Figure~\ref{chd} as an increase in  electron
density on the axis directed along the Al$_\mathrm{s}$-O$_\mathrm{m}$ bond 
and a depletion of electron density at O$_\mathrm{m}$. Calculating the electron transfer
from the electron density difference $\Delta n(\mathbf{r})$ we find that roughly 0.05
electrons are transferred to the Al$_\mathrm{s}$-O$_\mathrm{m}$ 
bond from the lone-pair electrons on O$_\mathrm{m}$.

It is worth noting that the adsorption energy found here
for methanol is very similar to the adsorption energy for
water on $\alpha$-Al$_2$O$_3$(0001). Hass et al.\cite{hsca:jpcb:5527}
obtained a molecular adsorption energy of 1.01 eV/molecule (23.3 kcal/mol)
in the low-coverage regime by DFT calculations, as well as
the bond lengths reproduced in Table~\ref{tab:geometri}.
The similarity between water and methanol molecular adsorption
is not too surprising. The parts of methanol taking an
active part in the binding to the surface are the O$_\mathrm{m}$
and H$_\mathrm{m}$ atoms of the hydroxyl group, similar to the OH
group of adsorbed water. 

\section{Summary}
We have performed first-principles density-functional theory calculations
on the molecular adsorption of methanol at the $\alpha$-Al$_2$O$_3$(0001) surface.
With one methanol molecule adsorbed per surface aluminum atom the
adsorption energy is
calculated to be $E_\mathrm{ads}^g=1.03$ eV/molecule. With
molecule-molecule interactions
taken into account, the adsorption energy is slightly less,
$E_\mathrm{ads}=1.01$
eV/molecule. An electron density difference plot shows that methanol
adsorbs by
donating electron charge from the methanol oxygen to create a bond to the
surface aluminum. Upon adsorption of methanol, the surface atomic
structure changes. Most
notably, the spacing between the top Al and O layers change from 0.11
{\AA} to 0.33 {\AA}.

\section{Acknowledgments}
We thank C.~Ruberto for providing data facilitating the $\alpha$-Al$_2$O$_3$(0001) 
calculations. This work has been done in \textit{Light Metal Surface Science} a joint project
between SINTEF and NTNU, financed by The Norwegian Research Council, Hydro Aluminium,
Profillakkering AS, Norsk Industrilakkering AS, NORAL AS, Jotun Powder Coating AS, Electro
Vacuum AS, Dupont Powder Coatings and GSB. The work was supported by a grant for computing time 
from The Norwegian Research Council. {\O}.B.\ benefited from a mobility scholarship from NorFA.  
The work of E.S.~was supported by the Swedish Research 
Council (VR) and the Trygger Foundation.

\clearpage

\begin{table}[ht]
  \begin{center}
    \caption{Calculated adsorption energies and selected bond lengths and 
bonding angle for methanol adsorbed 
on $\alpha$-Al$_2$O$_3$(0001). Subscript `m' labels atoms in the molecule, and
`s' atoms at the surface. H$_\mathrm{m}$ is the hydrogen of the hydroxyl group, 
and  H$_\mathrm{Me}$ the methyl-group hydrogen nearest to the surface.
The bond lengths are compared to the low-coverage bond lengths of molecularly 
adsorbed water as found by DFT calculations in 
Ref.~\protect\citen{hsca:jpcb:5527}.\vspace{0.5cm}}
    \begin{tabular}{lccc}
     \hline\hline
& Gas phase & Adsorbed & Adsorbed \\
& methanol &  methanol & H$_2$O (Ref.~\citen{hsca:jpcb:5527}) \\
\hline
$E_\mathrm{ads}$ (eV/molecule) & - & 1.01 & -\\
$E^\mathrm{g}_\mathrm{ads}$ (eV/molecule) & - & 1.03 & 1.01 \\
$d$(O$_\mathrm{m}$--Al$_\mathrm{s}$) ({\AA}) & - & 1.995 & 1.953 \\
$d$(H$_\mathrm{m}$--O$_\mathrm{s}$) ({\AA}) & - & 1.735 & - \\
$d$(O$_\mathrm{m}$--H$_\mathrm{m}$) ({\AA})& 0.979 & 1.022 & 0.978\\
$d$(C$_\mathrm{m}$--O$_\mathrm{m}$) ({\AA}) & 1.426 & 1.441 & - \\
$d$(H$_\mathrm{Me}$--O$_\mathrm{s}$) ({\AA}) & - & 2.699 & - \\
$\angle$C$_\mathrm{m}$O$_\mathrm{m}$H$_\mathrm{m}$ (deg) & 108.7 & 111.3 & -\\
\hline\hline
    \end{tabular}
    \label{tab:geometri}
  \end{center}
\end{table}

\clearpage

\begin{figure}
\begin{center}
\scalebox{0.3}{\includegraphics{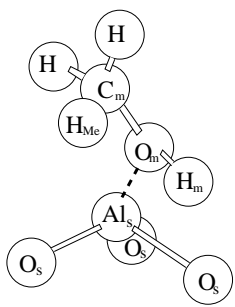}}
\caption{\label{adsmet}Ball-and-stick model of methanol adsorbed at the Al$_2$O$_3$ surface.
Only the outermost O and Al surface layers are shown. The bond between the methanol 
oxygen (O$_\mathrm{m}$) and the surface aluminum (Al$_\mathrm{s}$) is shown as a dashed
line. The atoms are named according to Table~\protect\ref{tab:geometri}. }
\end{center}
\end{figure}

\end{document}